\documentclass[cameraready]{Interspeech}

\title{RAT: Reference-Augmented Training for ASV Anti-Spoofing}

\author[orcid=0009-0000-5722-0571]{Vojtěch}{Staněk}
\author[orcid=0000-0002-4717-1910]{Anton}{Firc}
\author[orcid=0009-0004-6055-5136]{Jakub}{Reš}
\author[orcid=0000-0002-9009-2193]{Kamil}{Malinka}

\address{
    Security@FIT, Brno University of Technology, Czech Republic
}

\email{\{istanek, ifirc, iresj, malinka\}@fit.vut.cz}


\usepackage{comment}

\usepackage{booktabs}
\usepackage{multirow}
\usepackage{makecell}
\usepackage{svg}
\usepackage{adjustbox}
\usepackage{enumitem}

\begin{document} 

\maketitle

\begin{abstract}
    We introduce a spoofing countermeasure architecture conditioned on speaker-reference recordings, but observe that it converges to a solution that effectively ignores the reference during inference. Surprisingly, training with a reference channel induces invariance that improves deepfake detection, even when the reference is absent or mismatched during inference. Based on this observation, we propose a Reference-Augmented Training (RAT) strategy. RAT yields improved detection performance compared to single-utterance baselines, even when the reference recording is replaced with a zero vector at inference. Through rigorous analysis, we demonstrate that the optimization process rapidly diminishes the reference contributions, leading to inference largely independent of the reference channel. Using RAT, we achieve state-of-the-art 2.57\% EER and 0.074 minDCF on the ASVspoof 5 benchmark with a single detector, surpassing even large ensemble systems.
\end{abstract}

\section{Introduction}

Automatic Speaker Verification (ASV) systems face increasing threats from sophisticated spoofing attacks~\cite{diffuse}, including deepfakes created by text-to-speech synthesis and voice conversion~\cite{Firc2, MalinkaVoiceAssistants}. Traditional spoofing countermeasures (CMs) operate on single test utterances~\cite{Firc2025} without leveraging additional speaker-specific information that may be available during enrollment or verification~\cite{tdcf, FircSpectrogram}. While Spoofing-Aware Speaker Verification (SASV) leverages this enrollment data to jointly perform identity verification and spoofing detection~\cite{jung2022sasv}, standard CMs remain separate single-input systems.

Inspired by prior work in face morphing detection~\cite{Ibsen2021} and ASV anti-spoofing~\cite{liu23o_interspeech}, we intended to create a system enhanced by speaker-reference recordings. However, we observe a surprising phenomenon: using our architecture, training a spoofing detector with an additional reference recording input improves performance and generalization even when the reference is removed during inference. This suggests that reference-augmented training induces beneficial cues for the detector without enforcing strict reference dependence.

\textbf{Contributions.} Our contributions are as follows:
\begin{itemize}
    \item We propose Reference-Augmented Training (RAT), a training strategy that conditions on reference recordings via a reference-informed block (RIB) architecture.
    \item We demonstrate through functional and mechanistic analysis that reference influence diminishes during the training process, while yielding a model that retains its performance even after the reference channel is removed.
    \item We achieve state-of-the-art detection performance with RAT on ASVspoof 5 with 2.57\% EER and 0.074 minDCF, surpassing even large single-utterance fusions. We release the code, model weights, and the full-scored ASVspoof 5 evaluation at {\footnotesize \url{https://github.com/Security-FIT/RAT}}.
\end{itemize}

\section{Background}

\subsection{ASV spoofing countermeasures}

Current anti-spoofing systems utilize pretrained Self-supervised learning (SSL) models such as Wav2Vec2~\cite{xu24_asvspoof} or WavLM~\cite{stourbe24_asvspoof} due to their ability to extract rich speaker representations. These rich features are further processed and pooled, most commonly by Graph Attention Networks~\cite{Tak2021} from the AASIST framework~\cite{Jung2022aasist,xia24_asvspoof,scdf}. Beyond AASIST, architectures incorporate MHFA~\cite{rohdin24_asvspoof}, or Emphasized Channel Attention, Propagation, and Aggregation (ECAPA)~\cite{kulkarni24_asvspoof}. ResNet-based architectures are also used for both feature extraction and processing~\cite{dao24_asvspoof}. Additionally, combining multiple models~\cite{chan24_asvspoof, evolutionary_fusion} can significantly boost performance compared to using any of the classifiers separately. Despite advances in deepfake speech detection, current systems predominantly rely on data-driven architectures that do not incorporate methods that leverage prior knowledge or reference recordings.

\subsection{Reference-based methods}

While standard CMs operate as single-input binary classifiers~\cite{Firc3}, reference-based methods leverage enrollment data available in ASV scenarios to detect anomalies~\cite{tdcf}. This paradigm is central to SASV~\cite{jung2022sasv, wang24_asvspoof}, where common approaches fuse ASV scores with CM scores under the hypothesis that spoofing attacks may yield distinct score distributions compared to genuine target trials~\cite{rohdin24_asvspoof, wang24l_interspeech}. More recent deep learning approaches employ Siamese networks or attention mechanisms to explicitly compare the test utterance against a reference recording~\cite{liu23o_interspeech}. The premise is that while deepfakes may mimic high-level speaker identity, they fail to replicate fine-grained characteristics consistent with the enrollment recordings. These architectures rely on reference samples, but, as we discovered, these reference samples are not necessary during countermeasure inference to maintain the learned benefits.

\section{Architecture \& Methodology}

\begin{figure*}
    \centering
    \includegraphics[width=0.71\linewidth]{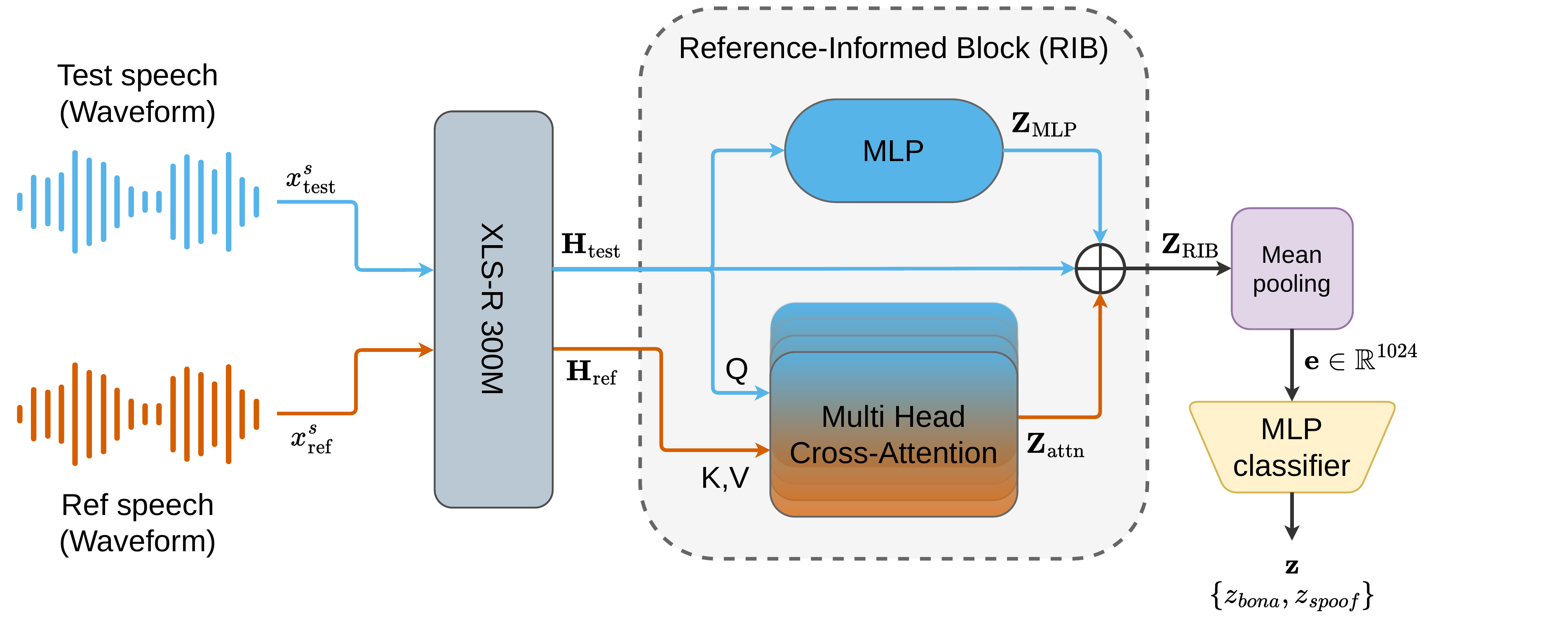}
    \caption{Proposed RAT architecture with the Reference-Informed Block (RIB), mean pooling, and a downstream MLP classifier.}
    \label{fig:architecture}
    \vspace{-0.5em}
\end{figure*}

Our architecture consists of three main components: an SSL feature extractor, a reference-informed block, and a downstream classifier, presented in \autoref{fig:architecture}.

    \textbf{SSL Feature Extractor}: We use a pre-trained XLS-R~\cite{Babu2021} model based on the Wav2Vec2 architecture with 300M parameters as our base feature extractor, shared for both the reference and test recordings. For each time frame, it extracts a 1024-dimensional feature vector. We utilize features from all 24 transformer layers of XLS-R, i.e., 
    \begin{align*}
    \textbf{H}_{\text{ref}} &= \text{XLS-R}(x_{\text{ref}}^s) \in \mathbb{R}^{24 \times t_{\text{ref}} \times 1024} \\ 
    \textbf{H}_{\text{test}} &= \text{XLS-R}(x_{\text{test}}^s) \in \mathbb{R}^{24 \times t_{\text{test}} \times 1024}
    \end{align*}
    are the extracted representations, where $x_{\text{ref}}^s, x_{\text{test}}^s$ are the reference and test recording from the same speaker $s$ with $t_{\text{ref}}$ and $t_{\text{test}}$ as the temporal dimensions of the reference and test recordings, respectively.
    
    \textbf{Reference-Informed Block (RIB)}: Before processing, we normalize the extracted features $\textbf{H}_{\text{ref}}, \textbf{H}_{\text{test}}$ using Layer Normalization (LN). Consequently, RIB processes embeddings through two parallel branches:
    
    \begin{enumerate}
        \item \textit{MLP Branch}: A lightweight multi-layer perceptron processes only test embeddings. It consists of two linear layers with ReLU nonlinearity and a 4× hidden expansion (1024→4096→1024) applied to each test embedding.
        \begin{equation*}
        \mathbf{Z}_{\text{MLP}} = \text{MLP}(\mathbf{H}_{\text{test}})
        \end{equation*}
        
        \item \textit{Cross-Attention Branch}: Multi-head cross-attention allows test embeddings to attend to reference embeddings. The test embeddings $\mathbf{H}_{\text{test}}$ serve as queries, while the reference embeddings $\mathbf{H}_{\text{ref}}$ act as keys and values with learnable parameters as usual~\cite{attention}:
        \begin{equation*}
        \mathbf{Z}_{\text{attn}} = \text{MultiHeadAttn}(\mathbf{H}_{\text{test}}, \mathbf{H}_{\text{ref}}, \mathbf{H}_{\text{ref}})
        \end{equation*}
    \end{enumerate}
    
    Based on preliminary experiments, we used 4 attention heads in our model. The outputs are combined with a residual connection:
    \begin{equation*}
        \mathbf{Z}_{\text{RIB}} = \mathbf{Z}_{\text{MLP}} + \mathbf{Z}_{\text{attn}} + \mathbf{H}_{\text{test}}
    \end{equation*}  
    
    This residual design ensures that when reference information is weak or unavailable (e.g., $x_{\text{ref}}^s = \mathbf{0}$), the model can still rely on test-only pathways. Finally, we normalize the output using another LN.
    
    \textbf{Pooling and Classification}: We apply mean pooling across the 24 SSL layers and the temporal dimension, which results in a single embedding $\mathbf{e} \in \mathbb{R}^{1024}$.
    The embedding is passed to a simple 3-layer MLP classifier with ReLU activations. The output is a 2D vector $\mathbf{z} =  \{z_{\text{bona}}, z_{\text{spoof}}\}$, i.e., logits corresponding to the bona fide and spoof classes. For evaluation, we use the $z_{\text{bona}}$ logit as the final spoofing countermeasure score.

\subsection{Reference sampling (pair construction)}

The dataloader provides pairs of samples from the same speaker. For each test recording, a pool of samples from the same speaker is gathered to randomly choose a bona fide reference recording. Formally, let $\mathcal{U}$ be the set of recordings in a dataset and $\mathcal{S}$ be the set of speakers in $\mathcal{U}$. For each speaker $s \in \mathcal{S}$, let $\mathcal{U}_s$ denote all recordings of $s$ and $\mathcal{B}_s \subseteq \mathcal{U}_s$ the subset of bona fide recordings of speaker $s$. For each test recording $x_{\text{test}}^{s} \in \mathcal{U}_s$, we create a reference pool of recordings $\mathcal{R}_s = \mathcal{B}_s \setminus \{x_{\text{test}}^s\}$ and sample a random reference recording $x_{\text{ref}}^s \sim \mathrm{Unif}(\mathcal{R}_s)$. The dataloader yields ordered pairs $(x_{\mathrm{ref}}^s, x_{\mathrm{test}}^s)$ such that both items come from the same speaker and the reference is bona fide, serving as input for the model.

This procedure is used during training and evaluation, unless stated otherwise, with samples drawn only from their respective subsets (i.e., the same train/dev/eval subset). Recordings are used as-is, i.e., without truncation or auxiliary padding\footnote{Within each mini-batch, we zero-pad all waveforms to the length of the longest recording in that batch; no other padding is used.}, enabling the model to process variable-length inputs.

\subsection{Training procedure}

We train with a simple Cross-entropy loss function using the Adam optimizer. For full reproducibility, our training and evaluation framework, model weights, and the scored ASVspoof 5 evaluation trials are publicly available in a GitHub repository\footnote{\url{https://github.com/Security-FIT/RAT}}. We employ a two-stage training strategy. 

In the first stage, the pretrained XLS-R frontend is frozen, while all other components are trained together. This allows rapid adaptation to the anti-spoofing task while preserving SSL representations. The learning rate used in this stage is $10^{-3}$, with a batch size of $16$. In the second stage, we unfreeze and fine-tune the \mbox{XLS-R} frontend jointly with the rest of the components using a small batch size of $6$ and a learning rate of $10^{-6}$. 

We train for 5 epochs in Stage~1 (with the frontend frozen) and for 6 epochs in Stage~2 (joint fine-tuning). We do not use any special learning-rate scheduling or weight decay. The model with the lowest Equal Error Rate (EER) on the dev subset is selected for the final evaluation. 

\subsection{Dataset and Evaluation}

We use the recent ASVspoof 5 benchmark~\cite{wang24_asvspoof}, which includes diverse spoofing attacks, multiple acoustic conditions and recording devices, various codec degradations, but also adversarial attacks Malafide~\cite{malafide} and Malacopula~\cite{malacopula}. 
As usual, we use the training split for model training, the development split for selecting the best-performing model and analysis, and the evaluation set for final evaluation. The splits are disjoint by attack and speaker.
We evaluate in the Open condition for Track 1 (spoofing detection), as we're using a pretrained XLS-R model. While XLS-R would not be allowed as an official ASVspoof 5 submission, we isolate the effect of RAT under a fixed frontend, training protocol, and data. Results similar to our XLS-R baseline have been reported with ASVspoof5-allowed SSL models such as WavLM~\cite{peng2025hybridpruning, best_single_asvspoof5}, signaling that our baseline is directly comparable.

To improve robustness to reference variations, we augment both test and reference recordings during training. We applied stochastic data augmentation with 30\% chances of applying each of the following: time masking, mu-law encoding/decoding, RawBoost (LnL-ISD)~\cite{RawBoost}, applying one of several noise types (color, Gaussian, or Gaussian SNR), and one of several filter types (band-pass, high/low-pass, shelf, peaking). This augmentation strategy is crucial for RAT, as it prevents the model from over-relying on clean references.

To measure achieved performance, we use the standard Equal Error Rate (EER) and minimum Detection Cost Function (minDCF)~\cite{wang24_asvspoof}. We follow the evaluation protocol of ASVspoof~5 and use the same parameters for computing minDCF: $C_{\text{fa}} = 10, C_{\text{miss}} = 1, p_{\text{spf}} = 0.05$, where $C_{\text{fa}}$ is the cost of false acceptance, $C_{\text{miss}}$ is the cost of a miss (i.e., rejecting bona fide), and $p_{\text{spf}}$ is prior for the spoof class. The parameters reflect the assumption that bona fide utterances are (generally) far more common in practice (low $p_{\text{spf}}$), but when deepfake utterances are encountered and not detected, the cost is high.

\section{Results}

We first evaluate the detection performance of the proposed Reference-Augmented Training (RAT) strategy against single-utterance baselines. We compare our method (\textit{$\approx$328M} parameters) with the best single system reported in the available literature: WavLM + Hybrid Pruning (\textit{$\approx$86M})~\cite{peng2025hybridpruning}, as well as best models from the ASVspoof 5 challenge: best single system WavLM-SLIM (\textit{$\approx$101M})~\cite{best_single_asvspoof5} and the ASVspoof 5 winner \textit{T43}~\cite{chen24_asvspoof}, which is a fusion of 12 large models (est.~over \textit{3.5B}). 

We also compare against our standard single-input XLS-R-based baseline with mean pooling (\textit{$\approx$316M}) following the same training and evaluation protocol as RAT without the reference augmentation training, alongside an identical architecture that uses only the test embedding as input to the cross-attention block, effectively degrading it to self-attention (\textit{$\approx$328M}).

Table~\ref{tab:main_results} shows that RAT substantially improves over our single-utterance XLS-R baseline under an otherwise identical training recipe. Notably, the model trained with paired references remains strong even when evaluated with a zero-vector reference $x_{\text{ref}}^i = \mathbf{0}$, indicating that including a reference recording primarily benefits training dynamics and is not necessary during inference. The test-only control (i.e., the self-attention variant) improves over unprocessed mean pooling but remains clearly behind RAT, suggesting the gains are not explained by the architecture alone.

\begin{table}[htbp]
    \centering
    \caption{Results on ASVspoof 5 eval set. The \textit{T43} winner is a 12-model fusion~\cite{chen24_asvspoof}. External systems are reported from their respective papers; our controlled comparisons (with 95\%~CI*) are the XLS-R baselines trained with the same recipe.} 
    \label{tab:main_results}
    \begin{adjustbox}{max width=\linewidth}
    \begin{tabular}{@{}lll@{}}
        \toprule
        \textbf{Model} & \textbf{EER} & \textbf{minDCF} \\
        \midrule
        WavLM-SLIM \cite{best_single_asvspoof5} & 5.56\% {\scriptsize [N/A]} & 0.149 {\scriptsize [N/A]} \\
        WavLM + Hybrid Pruning \cite{peng2025hybridpruning} &  3.75\% {\scriptsize [N/A]} & 0.103 {\scriptsize [N/A]} \\
        \textit{T43} (ASVspoof 5 winner) \cite{chen24_asvspoof} & 2.59\% {\scriptsize [N/A]} & 0.075 {\scriptsize [N/A]} \\
        \midrule
        XLS-R + mean pooling (ours) & 4.87\% {\scriptsize [4.79, 4.93]} & 0.141 {\scriptsize [0.139, 0.143]} \\
        XLS-R + RAT (test only) & 3.58\% {\scriptsize [3.52, 3.65]} & 0.104 {\scriptsize [0.102, 0.105]} \\
        XLS-R + RAT (with reference) & 2.63\% {\scriptsize [2.59, 2.68]} & 0.075 {\scriptsize [0.073, 0.076]} \\
        \textbf{XLS-R + RAT (zero reference)}$^\dagger$ & \textbf{2.57\% {\scriptsize [2.52, 2.62]}} & \textbf{0.074 {\scriptsize [0.072, 0.075]}} \\
        \bottomrule
    \end{tabular}
    \end{adjustbox}
    \vspace{0pt}
    \raggedright\\
    \footnotesize{\emph{*Confidence intervals computed by bootstrapping (1000 runs).}}\\
    \footnotesize{\emph{$^\dagger$Trained with reference, evaluated with zero-vector reference.}}
\end{table}

Furthermore, to validate our hypothesis that reference conditioning primarily improves learning rather than being required at inference time, we conduct inference-time ablation of the reference recording:
\begin{enumerate}[topsep=0pt, itemsep=0pt]
    \item \textbf{Paired inference (default RAT)}: bona fide reference from the same speaker.
    \item \textbf{Additive noise:} added noise at 10\,dB and 20\,dB SNR.
    \item \textbf{Truncation:} reference cropped to 1\,s and 3\,s.
    \item \textbf{Silence:} reference input is replaced by zeros ($x_{\text{ref}}^i = \mathbf{0}$).
    \item \textbf{Noise-only:} reference replaced by Gaussian noise only.
    \item \textbf{Mismatched:} bona fide reference from a different speaker.
\end{enumerate}


Table~\ref{tab:degradations} shows that performance is stable under moderate noise and truncation. Even under strong perturbations, i.e., silence, noise-only, and speaker-mismatch, the performance degrades only very slightly. This shows that dependence on the reference is limited during evaluation. Using a zero or noise-only reference completely removes speaker information from the reference channel during inference, effectively reducing the RAT model to a single-utterance detector. The observed limited drop in performance demonstrates a graceful disconnection, suggesting the reference is no longer required at test time.

\begin{table}[htbp]
    \centering
    \caption{RAT performance (with 95\%~CI*) evaluated under various reference signal degradations applied during inference.}
    \label{tab:degradations}
    \begin{adjustbox}{max width=\linewidth}
    \begin{tabular}{@{}lll@{}}
        \toprule
        \textbf{Degradation} & \textbf{EER} & \textbf{minDCF} \\
        \midrule
        None (paired inference) & 2.63\% {\scriptsize [2.59, 2.68]} & 0.075 {\scriptsize [0.073, 0.076]} \\
        Additive Noise (10\,dB) & 2.63\% {\scriptsize [2.58, 2.66]} & 0.075 {\scriptsize [0.073, 0.076]} \\
        Additive Noise (20\,dB) & 2.64\% {\scriptsize [2.58, 2.68]} & 0.075 {\scriptsize [0.073, 0.076]} \\
        Truncation (1\,s) & 2.65\% {\scriptsize [2.59, 2.68]} & 0.075 {\scriptsize [0.073, 0.076]} \\
        Truncation (3\,s) & 2.63\% {\scriptsize [2.58, 2.66]} & 0.074 {\scriptsize [0.073, 0.075]} \\
        Silence & 2.57\% {\scriptsize [2.52, 2.62]} & 0.074 {\scriptsize [0.072, 0.075]} \\
        Noise Only & 2.68\% {\scriptsize [2.63, 2.73]} & 0.077 {\scriptsize [0.075, 0.078]} \\
        Mismatched & 2.63\% {\scriptsize [2.60, 2.69]} & 0.075 {\scriptsize [0.074, 0.077]} \\
        \bottomrule
    \end{tabular}
    \end{adjustbox}
    \vspace{0pt}
    \raggedright\\
    \footnotesize{\emph{*Confidence intervals computed by bootstrapping (1000 runs).}}
\end{table}

\section{Analysis of Training Dynamics}

\begin{figure*}[htbp]
    \centering
    \includegraphics[width=\linewidth]{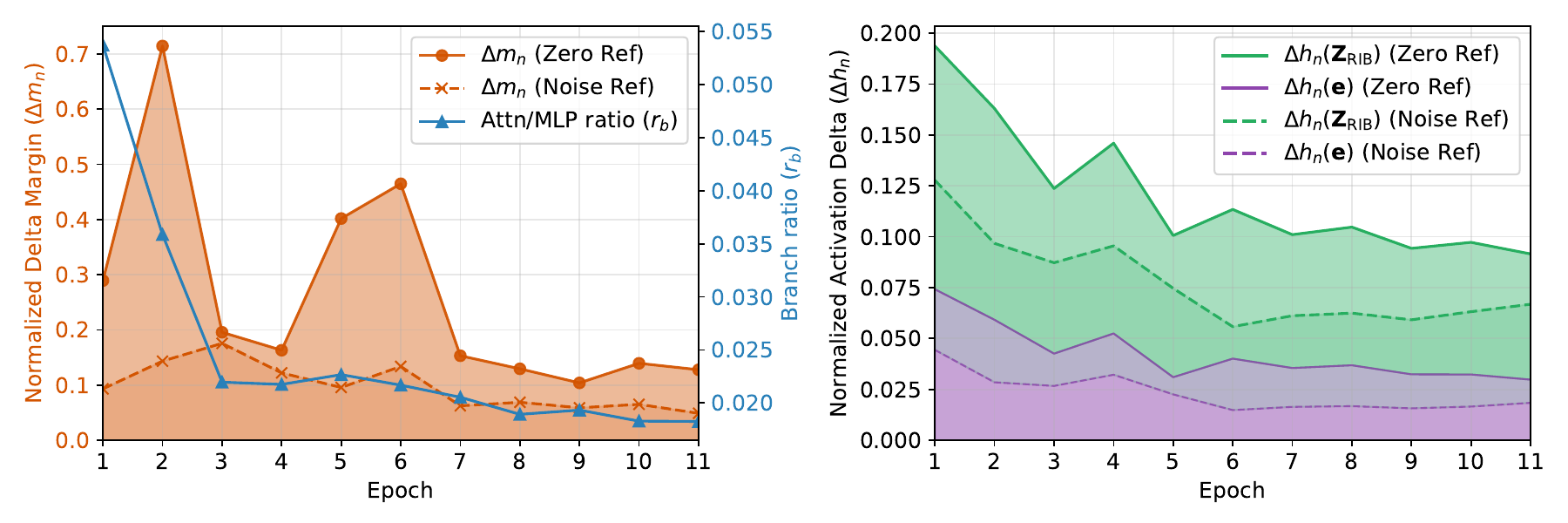}
    \caption{Training dynamics under zero and noise reference substitutions. \textbf{Left:} Normalized delta margin ($\Delta m_n$) and the attention-to-MLP branch ratio ($r_b$). \textbf{Right:} Normalized activation deltas ($\Delta h_n$) for the RIB block ($\mathbf{Z}_{\text{RIB}}$) and pooled embeddings ($\mathbf{e}$).}
    \label{fig:deltas}
\end{figure*}

To understand \textit{how} the model converges to this reference-invariant solution, we analyzed the model's and the RIB block's internal dynamics during training. To measure the reference reliance, we replace the reference $x_{\text{ref}}^i$ with a perturbed reference $\tilde{r}^i$ during evaluation after each epoch. We explored two options of $\tilde{r}^i$, that is, substituting either silence (i.e., zero-vector) or energy-matched\footnote{The energy of $\tilde{r}^i$ matches the energy of $x_{\text{ref}}^i$ to isolate the impact of missing speaker information from mere changes in signal amplitude.} noise vector as the reference input during evaluation. The measures are presented in \autoref{fig:deltas} and described below.

\subsection{Functional Dependence on Reference}

First, we quantify the functional impact of the reference input on the model's output. We calculate a margin:
\begin{equation*}
m(x_{\text{ref}}^i, x_{\text{test}}^i) = z_{\text{bona}}(x_{\text{ref}}^i, x_{\text{test}}^i) - z_{\text{spoof}}(x_{\text{ref}}^i, x_{\text{test}}^i)
\end{equation*}
as the difference between the bona fide and spoof logits. Intuitively, a larger margin means the model is more confident that the test utterance $x_{\text{test}}^i$ is bona fide, while a negative margin means the model is more confident in spoof prediction. We compute the normalized delta margin
\begin{equation*}
    \Delta m_n = \frac{ \mathbb{E}_i\left[ | m(x_{\text{ref}}^i, x_{\text{test}}^i) - m(\tilde{r}^i, x_{\text{test}}^i) | \right]}{\mathbb{E}_i\left[ | m(x_{\text{ref}}^i, x_{\text{test}}^i)| \right] }
\end{equation*}
which reflects the relative change in the margin (i.e., the decision statistic) when switching from $x_{\text{ref}}^i$ to $\tilde{r}^i$. 

As shown in \autoref{fig:deltas} (left), $\Delta m_n$ fluctuates during early training but steadily decays as the model converges. By the final epoch, replacing the bona fide reference with energy-matched noise alters the output margin by less than 5\%. While the reference may influence early learning, the final decision statistic becomes highly invariant to the reference signal.

\subsection{Mechanism analysis}

To understand the inner mechanism of RIB, we analyze two characteristics of our model.

\textbf{Branch Contribution Ratio.} We tracked the magnitude ratio of outputs of the attention branch and the MLP branch: 
\begin{equation*}
    r_{b} = \frac{||\mathbf{Z}_{\text{attn}}||_2}{||\mathbf{Z}_{\text{MLP}}||_2}
\end{equation*}
We observed that the model quickly learns to focus on the test embeddings. In the first epoch, the MLP branch already heavily dominates the cross-attention branch ($r_b \approx 0.05$). As training progresses, this ratio continues to decay to $r_b \approx 0.02$. This indicates that the optimization process naturally prioritizes the test-only features as the driving factor in the final decision.

\textbf{Tracing Change in Activations.} To localize where reference information influences the network, we measure the change in activations of model components when the reference input is replaced by $\tilde{r}^i$ while keeping $x_{\text{test}}^i$ fixed. Let $\mathbf{Z}^i$ denote an internal representation computed using the bona fide reference $x_{\text{ref}}^i$, and let $\mathbf{\tilde{Z}}^i$ denote the same representation computed using the perturbed reference $\tilde{r}^i$. We compute a normalized activation delta:
\begin{equation*}
    \Delta h_n = \mathbb{E}_i \left[ \frac{||\mathbf{Z}^i - \mathbf{\tilde{Z}}^i||_2}{||\mathbf{Z}^i||_2} \right]
\end{equation*}
We compute $\Delta h_n$ for $\mathbf{Z}^i \in \{ \mathbf{Z}_{\text{attn}}, \mathbf{Z}_{\text{RIB}}, \mathbf{e} \}$ as defined earlier. This \textit{delta trace} provides a component-wise diagnostic of dependence on reference. Large $\Delta h_n$ indicates that swapping the reference significantly changes the internal representation at that point, and small $\Delta h_n$ indicates effective invariance to reference input. 

As expected, the cross-attention output $\mathbf{Z}_{\text{attn}}$ is most sensitive to reference perturbations ($\Delta h_n \in [0.7, 1.5] $), confirming that the reference signal is included. 
The sensitivity drops significantly after the fusion stage $\mathbf{Z}_{\text{RIB}}$, and it is further diminished by the subsequent components in $\mathbf{e}$ as visible in \autoref{fig:deltas} (right). Despite this invariance, the logits and margin can still slightly change under reference replacement, especially with highly out-of-distribution references such as zero-vector silence, indicating that reference information serves as a somewhat bounded corrective signal rather than an important dependency.


\section{Conclusion}

We proposed Reference-Augmented Training (RAT), a training paradigm for ASV anti-spoofing that conditions the model on reference speaker recordings. By employing a Reference-Informed Block (RIB) with cross-attention, we discovered that the network utilizes the reference during early training as a corrective signal to better isolate spoofing artifacts, but eventually converges to a reference-invariant solution. Our analysis revealed that the optimization process gradually diminishes the reference's influence, gracefully disconnecting the reference channel. Ultimately, RAT achieves robust, state-of-the-art detection performance on the ASVspoof~5 benchmark, with 2.57\% EER and 0.074 minDCF, surpassing even large multi-model fusion systems with less than 10\% of their parameters. Crucially, this performance is maintained even when the reference signal is degraded, mismatched, or entirely absent during inference, allowing it to operate effectively as a conventional single-utterance detector.

\section{Acknowledgments}

This work was supported by the Brno University of Technology internal project FIT-S-26-9011. Computational resources were provided by the e-INFRA CZ project (ID:90254), supported by the Ministry of Education, Youth and Sports of the Czech Republic.

\section{Generative AI Use Disclosure}

During the preparation of this work, the authors used Generative AI Models (specifically Google Gemini, ChatGPT, and Grammarly) for language editing and text refinement. The authors reviewed and edited the output as needed and take full responsibility for the publication's content.



\bibliographystyle{IEEEtran}
\bibliography{mybib}

@article{Firc3,
  title   = {Deepfakes as a threat to a speaker and facial recognition: An overview of tools and attack vectors},
  journal = {Heliyon},
  volume  = {9},
  number  = {4},
  edition = {1},
  pages   = {e15090},
  year    = {2023},
  issn    = {2405-8440},
  doi     = {10.1016/j.heliyon.2023.e15090},
  author  = {Anton Firc and Kamil Malinka and Petr Hanáček}
}

@inproceedings{Babu2021,
  title     = {{XLS-R: Self-supervised Cross-lingual Speech Representation Learning at Scale}},
  author    = {Arun Babu and Changhan Wang and Andros Tjandra and Kushal Lakhotia and Qiantong Xu and Naman Goyal and Kritika Singh and Patrick {von Platen} and Yatharth Saraf and Juan Pino and Alexei Baevski and Alexis Conneau and Michael Auli},
  year      = {2022},
  booktitle = {Interspeech 2022},
  pages     = {2278--2282},
  doi       = {10.21437/Interspeech.2022-143},
  issn      = {2958-1796},
}

@inproceedings{Ibsen2021,
  author       = {Ibsen, M. and Gonzalez-Soler, L. J. and Rathgeb, C. and Drozdowski, P. and Gomez-Barrero, M. and Busch, C.},
  booktitle    = {2021 IEEE International Workshop on Information Forensics and Security (WIFS)},
  title        = {{Differential Anomaly Detection for Facial Images}},
  year         = {2021},
  pages        = {1-6},
  doi          = {10.1109/WIFS53200.2021.9648392},
  organization = {IEEE},
  isbn         = {978-1-6654-1717-4}
}

@inproceedings{wang24_asvspoof,
  title     = {{ASVspoof 5: crowdsourced speech data, deepfakes, and adversarial attacks at scale}},
  author    = {Xin Wang and Héctor Delgado and Hemlata Tak and Jee-weon Jung and Hye-jin Shim and Massimiliano Todisco and Ivan Kukanov and Xuechen Liu and Md Sahidullah and Tomi H. Kinnunen and Nicholas Evans and Kong Aik Lee and Junichi Yamagishi},
  year      = {2024},
  booktitle = {{The Automatic Speaker Verification Spoofing Countermeasures Workshop (ASVspoof 2024)}},
  pages     = {1--8},
  doi       = {10.21437/ASVspoof.2024-1},
}

@inproceedings{Tak2021,
  author    = {Hemlata Tak and Jee-weon Jung and Jose Patino and Massimiliano Todisco and Nicholas Evans},
  title     = {{Graph Attention Networks for Anti-Spoofing}},
  year      = 2021,
  booktitle = {Proc. Interspeech 2021},
  pages     = {2356--2360},
  doi       = {10.21437/Interspeech.2021-993}
}

@INPROCEEDINGS{Jung2022aasist,
  author={Jung, Jee-weon and Heo, Hee-Soo and Tak, Hemlata and Shim, Hye-jin and Chung, Joon Son and Lee, Bong-Jin and Yu, Ha-Jin and Evans, Nicholas},
  booktitle={ICASSP 2022 - 2022 IEEE International Conference on Acoustics, Speech and Signal Processing (ICASSP)}, 
  title={{AASIST: Audio Anti-Spoofing Using Integrated Spectro-Temporal Graph Attention Networks}}, 
  year={2022},
  volume={},
  number={},
  pages={6367-6371},
  doi={10.1109/ICASSP43922.2022.9747766},
}

@inproceedings{xia24_asvspoof,
  title     = {{A single end-to-end voice anti-spoofing model with graph attention and feature aggregation for ASVspoof 5 Challenge}},
  author    = {Weijiang Xia and others},
  year      = {2024},
  booktitle = {The Automatic Speaker Verification Spoofing Countermeasures Workshop},
  doi       = {10.21437/ASVspoof.2024-18},
}

@inproceedings{rohdin24_asvspoof,
  title     = {{BUT systems and analyses for the ASVspoof 5 Challenge}},
  author    = {Johan Rohdin and Lin Zhang and Plchot Oldřich and Vojtěch Staněk and David Mihola and Junyi Peng and Themos Stafylakis and Dmitriy Beveraki and Anna Silnova and Jan Brukner and Lukáš Burget},
  year      = {2024},
  booktitle = {The Automatic Speaker Verification Spoofing Countermeasures Workshop (ASVspoof 2024)},
  pages     = {24--31},
  doi       = {10.21437/ASVspoof.2024-4},
}

@inproceedings{chan24_asvspoof,
  title     = {{Enhancing spoofing detection in ASVspoof 5 Workshop 2024: fusion of WavLM-ResNet18-SA for optimal performance against speech deepfakes}},
  author    = {Po-Cheng Chan and Wei-Yu Chen and Jia-Ching Wang},
  year      = {2024},
  booktitle = {The Automatic Speaker Verification Spoofing Countermeasures Workshop},
  pages     = {158--162},
  doi       = {10.21437/ASVspoof.2024-23},
}

@inproceedings{dao24_asvspoof,
  title     = {{ASVspoof 5 Challenge: advanced ResNet architectures for robust voice spoofing detection}},
  author    = {Anh-Tuan Dao and Mickael Rouvier and Driss Matrouf},
  year      = {2024},
  booktitle = {The Automatic Speaker Verification Spoofing Countermeasures Workshop},
  pages     = {163--169},
  doi       = {10.21437/ASVspoof.2024-24},
}

@inproceedings{stourbe24_asvspoof,
  title     = {{Exploring WavLM back-ends for speech spoofing and deepfake detection}},
  author    = {Théophile Stourbe and Victor Miara and Theo Lepage and Reda Dehak},
  year      = {2024},
  booktitle = {The Automatic Speaker Verification Spoofing Countermeasures Workshop},
  pages     = {72--78},
  doi       = {10.21437/ASVspoof.2024-11},
}

@inproceedings{xu24_asvspoof,
  title     = {{SZU-AFS antispoofing system for the ASVspoof 5 Challenge}},
  author    = {Yuxiong Xu and Jiafeng Zhong and Sengui Zheng and Zefeng Liu and Bin Li},
  year      = {2024},
  booktitle = {The Automatic Speaker Verification Spoofing Countermeasures Workshop},
  pages     = {64--71},
  doi       = {10.21437/ASVspoof.2024-10},
}

@inproceedings{kulkarni24_asvspoof,
  title     = {{Exploring generalization to unseen audio data for spoofing: insights from SSL models}},
  author    = {Atharva Kulkarni and Hoan My Tran and Ajinkya Kulkarni and Sandipana Dowerah and Damien Lolive and Mathew Maginai Doss},
  year      = {2024},
  booktitle = {The Automatic Speaker Verification Spoofing Countermeasures Workshop (ASVspoof 2024)},
  pages     = {86--93},
  doi       = {10.21437/ASVspoof.2024-13},
}

@INPROCEEDINGS{RawBoost,
  author={Tak, Hemlata and Kamble, Madhu and Patino, Jose and Todisco, Massimiliano and Evans, Nicholas},
  booktitle={ICASSP 2022 - 2022 IEEE International Conference on Acoustics, Speech and Signal Processing (ICASSP)}, 
  title={{Rawboost: A Raw Data Boosting and Augmentation Method Applied to Automatic Speaker Verification Anti-Spoofing}}, 
  year={2022},
  volume={},
  number={},
  pages={6382-6386},
  doi={10.1109/ICASSP43922.2022.9746213}}

@inproceedings{chen24_asvspoof,
  title     = {{USTC-KXDIGIT system description for ASVspoof5 Challenge}},
  author    = {Yihao Chen and Haochen Wu and Nan Jiang and Xiang Xia and Qing Gu and YunQi Hao and Pengfei Cai and Yu Guan and Jialong Wang and Wei-Lin Xie and Lei Fang and Sian Fang and Yan Song and Wu Guo and Lin Liu and Minqiang Xu},
  year      = {2024},
  booktitle = {The Automatic Speaker Verification Spoofing Countermeasures Workshop (ASVspoof 2024)},
  pages     = {109--115},
  doi       = {10.21437/ASVspoof.2024-16},
}

@incollection{scdf,
author = "Staněk, Vojtěch and Srna, Karel and Firc, Anton and Malinka, Kamil",
title = {{SCDF: A Speaker Characteristics Deepfake Speech Dataset for Bias Analysis}},
year = 2025,
doi = "10.18420/biosig_2025_005",
booktitle = "BIOSIG 2025",
publisher = "Gesellschaft für Informatik e.V.",
issn = "2944-7682",
}

@inproceedings{diffuse,
   title={{Diffuse or Confuse: A Diffusion Deepfake Speech Dataset}},
   DOI={10.1109/biosig61931.2024.10786752},
   booktitle={2024 International Conference of the Biometrics Special Interest Group (BIOSIG)},
   publisher={IEEE},
   author={Firc, Anton and Malinka, Kamil and Hanáček, Petr},
   year={2024},
   month=sep, pages={1–7}
}

@inproceedings{FircSpectrogram,
author = {Firc, Anton and Malinka, Kamil and Han\'{a}\v{c}ek, Petr},
title = {{Deepfake Speech Detection: A Spectrogram Analysis}},
year = {2024},
isbn = {9798400702433},
publisher = {Association for Computing Machinery},
address = {New York, NY, USA},
doi = {10.1145/3605098.3635911},
booktitle = {Proceedings of the 39th ACM/SIGAPP Symposium on Applied Computing},
pages = {1312–1320},
numpages = {9},
location = {Avila, Spain},
series = {SAC '24}
}

@inproceedings{Firc2,
author = {Firc, Anton and Malinka, Kamil},
title = {{The dawn of a text-dependent society: deepfakes as a threat to speech verification systems}},
year = {2022},
isbn = {9781450387132},
publisher = {Association for Computing Machinery},
address = {New York, NY, USA},
doi = {10.1145/3477314.3507013},
booktitle = {Proceedings of the 37th ACM/SIGAPP Symposium on Applied Computing},
pages = {1646–1655},
numpages = {10},
location = {Virtual Event},
series = {SAC '22}
}

@InProceedings{MalinkaVoiceAssistants,
author="Malinka, Kamil and others",
title={{Resilience of Voice Assistants to Synthetic Speech}},
booktitle="Computer Security -- ESORICS 2024",
year="2024",
publisher="Springer Nature Switzerland",
address="Cham",
pages="66--84",
isbn="978-3-031-70879-4"
}

@inproceedings{malafide,
  title     = {{Malafide: a novel adversarial convolutive noise attack against deepfake and spoofing detection systems}},
  author    = {Michele Panariello and Wanying Ge and Hemlata Tak and Massimiliano Todisco and Nicholas Evans},
  year      = {2023},
  booktitle = {{Interspeech 2023}},
  pages     = {2868--2872},
  doi       = {10.21437/Interspeech.2023-703},
  issn      = {2958-1796},
}

@inproceedings{malacopula,
  title     = {{Malacopula: adversarial automatic speaker verification attacks using a neural-based generalised Hammerstein model}},
  author    = {Massimiliano Todisco and Michele Panariello and Xin Wang and Héctor Delgado and Kong Aik Lee and Nicholas Evans},
  year      = {2024},
  booktitle = {{The Automatic Speaker Verification Spoofing Countermeasures Workshop (ASVspoof 2024)}},
  pages     = {94--100},
  doi       = {10.21437/ASVspoof.2024-14},
}

@inproceedings{attention,
 author = {Vaswani, Ashish and Shazeer, Noam and Parmar, Niki and Uszkoreit, Jakob and Jones, Llion and Gomez, Aidan N and Kaiser, Lukasz and Polosukhin, Illia},
 booktitle = {Advances in Neural Information Processing Systems},
 editor = {I. Guyon and U. Von Luxburg and S. Bengio and H. Wallach and R. Fergus and S. Vishwanathan and R. Garnett},
 pages = {},
 publisher = {Curran Associates, Inc.},
 title = {Attention is All you Need},
 volume = {30},
 year = {2017}
}

@inproceedings{liu23o_interspeech,
  title     = {{Speaker-Aware Anti-spoofing}},
  author    = {Xuechen Liu and Md Sahidullah and Kong Aik Lee and Tomi Kinnunen},
  year      = {2023},
  booktitle = {{Interspeech 2023}},
  pages     = {2498--2502},
  doi       = {10.21437/Interspeech.2023-1323},
  issn      = {2958-1796},
}

@ARTICLE{tdcf,
  author={Kinnunen, Tomi and Delgado, Héctor and Evans, Nicholas and Lee, Kong Aik and Vestman, Ville and Nautsch, Andreas and Todisco, Massimiliano and Wang, Xin and Sahidullah, Md and Yamagishi, Junichi and Reynolds, Douglas A.},
  journal={IEEE/ACM Transactions on Audio, Speech, and Language Processing}, 
  title={{Tandem Assessment of Spoofing Countermeasures and Automatic Speaker Verification: Fundamentals}}, 
  year={2020},
  volume={28},
  number={},
  pages={2195-2210},
  doi={10.1109/TASLP.2020.3009494}
}

@inproceedings{wang24l_interspeech,
  title     = {{Revisiting and Improving Scoring Fusion for Spoofing-aware Speaker Verification Using Compositional Data Analysis}},
  author    = {Xin Wang and Tomi Kinnunen and Kong Aik Lee and Paul-Gauthier Noé and Junichi Yamagishi},
  year      = {2024},
  booktitle = {{Interspeech 2024}},
  pages     = {1110--1114},
  doi       = {10.21437/Interspeech.2024-422},
  issn      = {2958-1796},
}

@inproceedings{best_single_asvspoof5,
  title     = {{Learn from real: reality defender's submission to ASVspoof5 Challenge}},
  author    = {Yi Zhu and Chirag Goel and Surya Koppisetti and Trang Tran and Ankur Kumar and Gaurav Bharaj},
  year      = {2024},
  booktitle = {{The Automatic Speaker Verification Spoofing Countermeasures Workshop (ASVspoof 2024)}},
  pages     = {116--123},
  doi       = {10.21437/ASVspoof.2024-17},
}

@misc{peng2025hybridpruning,
      title={Hybrid Pruning: In-Situ Compression of Self-Supervised Speech Models for Speaker Verification and Anti-Spoofing}, 
      author={Junyi Peng and Lin Zhang and Jiangyu Han and Oldřich Plchot and Johan Rohdin and Themos Stafylakis and Shuai Wang and Jan Černocký},
      year={2025},
      eprint={2508.16232},
      archivePrefix={arXiv},
      primaryClass={eess.AS},
      url={https://arxiv.org/abs/2508.16232}, 
}

@inproceedings{Jung2022sasv,
    author = {Jung, Jee-weon and Tak, Hemlata and Shim, Hye-jin and Heo, Hee-Soo and Lee, Bong-Jin and Chung, Soo-Whan and Yu, Ha-Jin and Evans, Nicholas and Kinnunen, Tomi},
    title = {{SASV 2022: The First Spoofing-Aware Speaker Verification Challenge}},
    booktitle = {Proc. Interspeech (submitted)},
    year = {2022},
}

@ARTICLE{Firc2025,
	author = {Firc, Anton and Malinka, Kamil and Hanáček, Petr},
	title = {{Evaluation framework for deepfake speech detection: a comparative study of state-of-the-art deepfake speech detectors}},
	year = {2025},
	journal = {Cybersecurity},
	volume = {8},
	number = {1},
	doi = {10.1186/s42400-024-00346-1},
}

@misc{evolutionary_fusion,
      title={{Evolutionary Multi-Objective Fusion of Deepfake Speech Detectors}}, 
      author={Vojtěch Staněk and Martin Perešíni and Lukáš Sekanina and Anton Firc and Kamil Malinka},
      year={2026},
      eprint={2604.01330},
      archivePrefix={arXiv},
      primaryClass={cs.SD},
      url={https://arxiv.org/abs/2604.01330}, 
}

\end{document}